\documentclass[twocolumn]{aastex63}
\usepackage{amsmath,amstext,amssymb}
\usepackage{xspace}
\usepackage{lipsum}
\usepackage{multirow}

\newcommand{\UTokyo}{Institute for Cosmic Ray Research, The University of Tokyo, 5-1-5 Kashiwanoha, Kashiwa, Chiba 277-8582, Japan}
\newcommand{\Cardiff}{Cardiff University, Cardiff CF24 3AA, UK}
\newcommand{\CMU}{McWilliams Center for Cosmology, Department of Physics, Carnegie Mellon University, Pittsburgh, PA 15213, USA}
\newcommand{\Caltech}{California Institute of Technology, Pasadena, CA 91125, USA}

\newcommand{\GW}{GW190425\xspace}
\newcommand{\FRB}{FRB20190425A\xspace}
\newcommand{\GAL}{UGC10667\xspace}
\newcommand{\low}{low-spin\xspace}
\newcommand{\high}{high-spin\xspace}

\shortauthors{Maga\~na~Hernandez, d'Emilio et al.}

\begin{document}

\title{On the association of \GW with its potential electromagnetic counterpart FRB 20190425A}
\shorttitle{\GW and \FRB Association}

\author[0000-0003-2362-0459]{Ignacio Maga\~na~Hernandez}
\affiliation{\CMU}
\email{imhernan@andrew.cmu.edu}

\author[0000-0001-6145-8187]{Virginia d’Emilio}
\affiliation{\Cardiff}
\affiliation{\Caltech}
\email{demiliov@caltech.edu}

\author[0000-0002-8445-6747]{Soichiro Morisaki}
\affiliation{\UTokyo}

\author[0000-0002-3615-3514]{Mohit Bhardwaj}

\affiliation{\CMU}

\author[0000-0002-6011-0530]{Antonella Palmese}
\affiliation{\CMU}

\begin{abstract}
Recent work by \cite{moroianu2022assessment} has suggested that the binary neutron star (BNS) merger \GW might have a potential fast radio burst (FRB) counterpart association, \FRB, at the 2.8$\sigma$ level of confidence with a likely host galaxy association, namely \GAL. The authors argue that the observations are consistent with a long-lived hypermassive neutron star (HMNS) that formed promptly after the BNS merger and was stable for approximately 2.5 hours before promptly collapsing into a black hole. Recently, \cite{2023arXiv231010018B} conclusively associated \FRB with \GAL, potentially providing a direct host galaxy candidate for \GW. In this work, we examine the multi-messenger association based on the space-time localization overlaps between \GW and the FRB host galaxy \GAL and find that the odds for a coincident association are $\mathcal{O}(5)$. We validate this estimate by using a Gaussian Process (GP) density estimator. Assuming that the association is indeed real, we then perform Bayesian parameter estimation on \GW assuming that the BNS event took place in \GAL.
We find that the viewing angle of \GW excludes an on-axis system at $p(\theta_v>30^o)\approx99.99$\%, highly favoring an off-axis system similar to GRB 170817A. 
We also find a slightly higher source frame total mass for the binary, namely, $m_{\rm{total}} = 3.42^{+0.34}_{-0.11} M_{\odot}$, leading to an increase on the probability of prompt collapse into a black hole and therefore disfavors the long-lived HMNS formation scenario. 
Given our findings, we conclude that the association between \GW and \FRB is disfavoured by current state-of-the-art gravitational-wave analyses.
\end{abstract}


\section{Introduction} \label{section:introduction}

The first detection of gravitational waves by Advanced LIGO \citep{aasi2015advanced} and Virgo \citep{acernese2014advanced} from the merger of two neutron stars, GW170817, allowed for the first multi-messenger studies using both gravitational wave data (GW) and electromagnetic observations (EM)~\citep{abbott2019properties}. Following the binary neutron star (BNS) merger, a burst of short gamma rays (sGRB) was detected by Fermi and INTEGRAL about 2 seconds after the GW emission~\citep{abbott2017gravitational}. As NS matter collided, a Kilonova (KN) was produced and was eventually observed 11.4 hours after GW170817 was detected~\citep{soares2017electromagnetic, Nicholl_2017}. This allowed for the unique identification of the host galaxy of GW170817, a relatively old and massive galaxy, namely NGC4993. 
Follow-up, radio observations determined a radio afterglow that was first observed around 100 days after GW170817 and that is still detectable to date~\citep{Balasubramanian:2022sie}. 

The detection of GW170817 and its many EM counterparts, in particular, GRB 170817A allowed for the study of the statistical significance that both the GW and sGRB events originated from a common astrophysical source. Given the two second time delay between GW170817 and GRB 170817A as well as the typical Fermi sGRB detection rate, it was concluded that from timing considerations alone, the coincident (common source) hypothesis was favored by $\mathcal{O}(10^6)$ times more than a mere chance of random association. Further studies (see for example \cite{Piotrzkowski:2021hhy}), used the small localization volume for GW170817 and its host galaxy, NGC4993, to study the statistical chance of spatial association. These studies arrived at similar conclusions, however, we note that the most stringent constraints were placed by the time delay between GW170817 and GRB 170817A. With such strong odds, it is believed that these two events and the follow-up EM observations were all due to the first detectable merger of neutron stars in both gravitational waves and light. 

During the first half of the LIGO-Virgo-KAGRA (LVK) third observing run (O3a)~\citep{LIGOScientific:2021usb, akutsu2021overview}, \GW, a second high-confidence detection of GWs from a BNS merger was discovered~\citep{LIGOScientific:2020aai}. The gravitational waves were observed initially only by the LIGO Livingston observatory but Virgo was also functional at the time. The two detector network detection did not allow for precise sky localization and \GW was localized to around $10^5 \ \rm{deg}^2$ in the sky. The large sky localization region and the fact that this event happened at a distance of about 200 Mpc did not allow for the detection of any confidently associated electromagnetic counterparts in low latency.

Recent work by \cite{moroianu2022assessment}, has found evidence at a 2.8$\sigma$ level, for the association between \GW and the fast radio burst \FRB detected by the CHIME Collaboration about 2.5 hours after the BNS merger with a likely host galaxy association, namely \GAL. 
Using approximately a couple orders of magnitude precise baseband localization of \FRB, \cite{2023arXiv231010018B} robustly associated \GAL as the host galaxy of \FRB, with a probability of chance association of $<$ 0.1\%.

Given the total mass for \GW was found to be $3.4^{+0.3}_{-0.1} \ M_{\odot}$, the LVK collaboration suggested that the two neutron stars most likely collapsed into a black hole promptly. This is consistent with our current constraints on the equation of state (EOS) for dense nuclear matter \citep{abbott2020gw190425}. \cite{moroianu2022assessment} and \cite{zhang2022physics}, however, put forward a proposed scenario to suggest a mechanism for the potential FRB emission and to explain the delay of the merger. 
In order for the proposed neutron star to not collapse directly into a black hole, it is necessary to invoke a highly spinning remnant, as this might provide increased mass support, as well as a stiffer EOS and potentially an exotic compact object as one of the binary components, e.g., a quark star. The hypermassive NS would then survive the direct collapse for about 2.5 hours until it collapses into a black hole ejecting its magnetosphere in the process leading to the production of \FRB. 

In this work, we re-examine the association between \GW and \FRB by considering spatial and temporal coincidence including the limited field of view of CHIME. We assume that \GAL is the host galaxy for both \GW and \FRB in our posterior odds calculations. We then perform GW parameter estimation on \GW under the assumption that \GAL is indeed the host for both transients to have a direct measurement of the viewing angle to the BNS event as well as improved mass estimates. 

\section{\GW and \FRB association} \label{section:association}

To examine the association between the gravitational-wave event \GW with its potential EM counterpart \FRB, we follow the formalism in \cite{ashton2018coincident,ashton2021current} to compute the posterior odds for a common source for the two transients. 
We compare two hypotheses: a common source $C$, in which the \GW post-merger remnant produces an FRB counterpart by ejecting its magnetosphere before collapsing onto a black hole \citep{moroianu2022assessment}; and a random coincidence hypothesis $R$, in which both events are entirely distinct.

The agreement between posterior distributions, under the common source hypothesis, is quantified by the posterior overlap integral. For a given set of parameters $\theta$, the integral is defined as,
\begin{equation}
    \mathcal{I}_{\theta} = \int \frac{p(\theta | d_{\textrm{GW}}, C)p(\theta | d_{\textrm{EM}}, C)}{\pi(\theta|C)}d\theta .
\end{equation}
where $p(\theta | d_{\textrm{GW}}, C)$ is the parameter's posterior given the GW observation, $p(\theta | d_{\textrm{EM}}, C)$ is the parameter's probability from EM and $\pi(\theta|C)$ is the parameter's prior distribution.

Considering both spatial and temporal coincidences between the GW and FRB observations, as mathematically derived in \cite{ashton2018coincident} Eq. (5), the posterior odds between the two competing hypotheses can then be calculated as,
\begin{equation}
\mathcal{O}_{C/R}=\pi_{C/R}\mathcal{I}_{D_L,\Omega}\mathcal{I}_{t_c}\approx \pi_{C/R}\mathcal{I}_{D_L}\mathcal{I}_{\Omega}\mathcal{I}_{t_c}
\end{equation}
where $\mathcal{I}_{D_L,\Omega}$ is the overlap integral for the 3-dimensional localization volumes between the two transients and $\mathcal{I}_{D_L}$ and $\mathcal{I}_{\Omega}$ are the overlap integrals for the approximately disjoint luminosity distance and sky localizations respectively. The temporal overlap integral is given by $\mathcal{I}_{t_c}$ and $\pi_{C/R}$ is the ratio of probabilities for the two hypotheses based solely on prior information, e.g., the detection rates for the transients.

\subsection{Spatial overlap}
To measure the posterior odds of \GW being associated with \FRB we use the publicly available LVK posterior samples on the parameters of \GW \citep{LIGOScientific:2020aai,LIGOScientific:2021djp},\citep{pe-data-doi-lvk}. 

\begin{figure}[h!]
   \centering
    \includegraphics[width=0.5\textwidth]{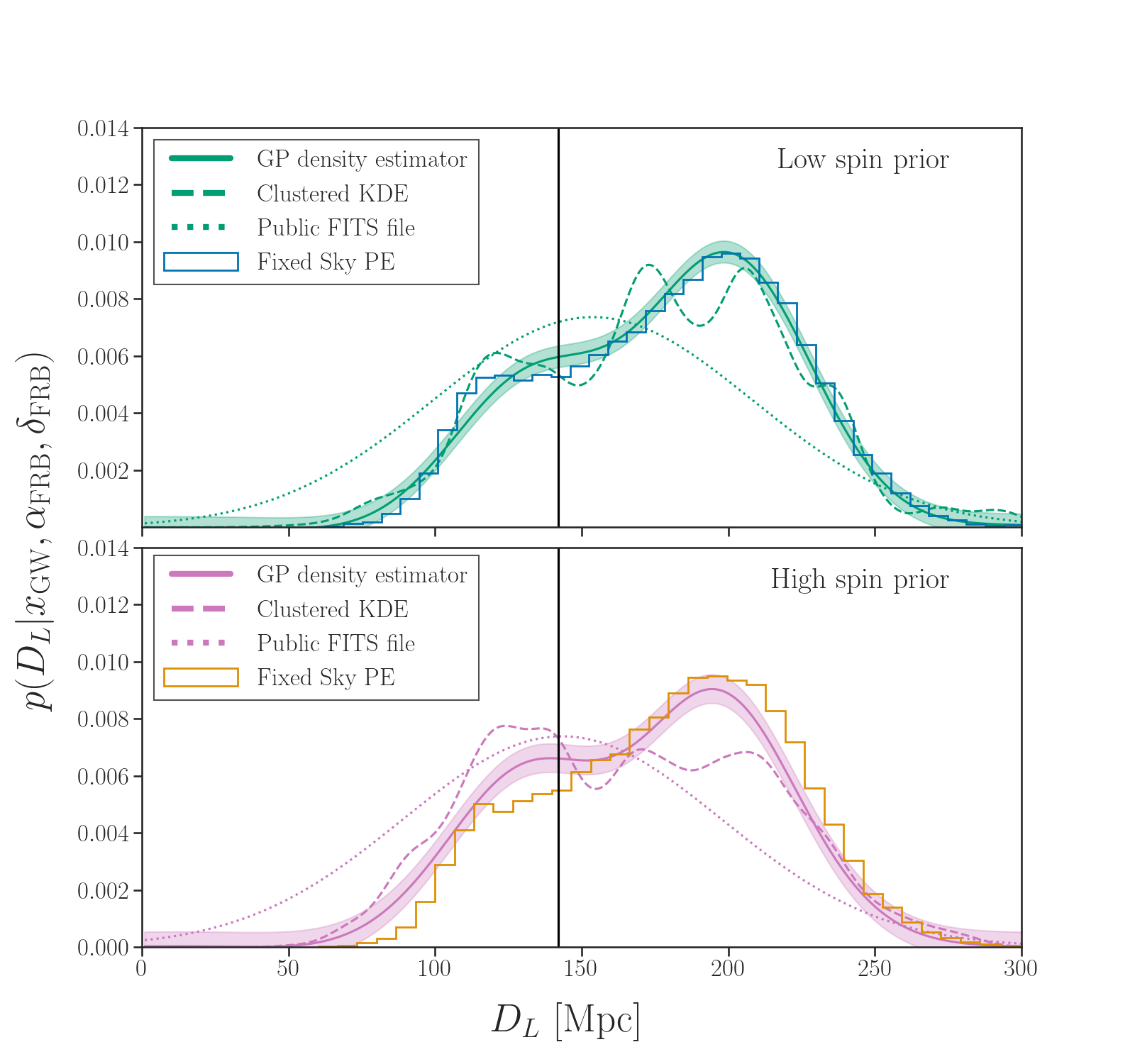}
   \caption{Posterior probability of the luminosity distance from GWTC-3 samples marginalised along the FRB line-of-sight (LOS) (black line). The top panel shows \low results, bottom panel shows \high results. The marginal posterior distribution is computed using localization \texttt{FITS} files (dotted line), a \textit{Clustered KDE} (dashed line), and a GP density estimator (solid line). The shaded band shows the GP's $2\sigma$ uncertainty. The LOS PE samples obtained in this work are also shown here for comparison (as histograms). }
   \label{fig:dl_cosmo}
\end{figure}

The joint posterior overlap integral $\mathcal{I}_{\rm{D_L,\Omega}}$ requires interpolating the three-dimensional posterior density $p(D_L,\Omega | d_{\textrm{GW}},C)$. Since \GW was not a well-localized event, the density surface presents degenerate correlations and non-Gaussianities. To assess the goodness of the three-dimensional fit, we look at the one-dimensional slice of the interpolation and of the GWTC-3 samples over the FRB sky location, as shown in Figure~\ref{fig:dl_cosmo}. For comparison purposes, we compute the interpolation with \texttt{ligo.skymap}'s \textit{ClusteredKDE}~\citep{singer2020} as well as with the publicly avaliable LVK 3D skymap. We also interpolate the posterior distribution with a Gaussian Process (GP) density estimator~\citep{d2021density}, which comes with an associated fit uncertainty. 
As best illustrated by Figure~\ref{fig:dl_cosmo}, we believe the public 3D skymap interpolation to be inaccurate in three dimensions since important density features are smoothed out. Hence we only report results obtained with the \textit{ClusteredKDE} and the GP density estimates.

To better understand the individual contributions of the joint integral, we also calculate the odds association by approximating $\mathcal{I}_{D_L,\Omega}\approx\mathcal{I}_{\rm{D_L}}\mathcal{I}_{\Omega}$.
Since the $\mathcal{I}_{\rm{D_L}}$ integral is in one-dimension, it is computed with a Gaussian KDE; while the two-dimensional $\mathcal{I}_{\Omega}$ is computed with the \texttt{ligo.skymap} package. We also compute both integral interpolations with a GP density estimator, for comparison. The latter are shown in Figure~\ref{fig:skymap-GP}.

\subsection{Instantaneous field-of-view of CHIME}
In Earth-fixed coordinates, CHIME looks directly over LIGO Hanford and near the region of the largest antenna response. Moreover, the LIGO-Virgo detector network preferentially detects signals from directly overhead/underneath. Due to this, FRBs observed within a few hours of a GW trigger will have a higher probability of chance sky position overlap than FRBs observed at other times. 

Therefore, we need to account for this non-negligible correlation between CHIME and the LIGO detectors due to CHIME's instantaneous field-of-view (FOV). 
We encode the correlations between the two instruments entirely in the spatial overlap prior, i.e. the denominator of Eq. (1), such that it corresponds to their common FOV viewing window.
We modify the default full sky prior $\pi(\Omega|\rm{Full \ Sky})$, from \citep{Romero-Shaw:2020owr}, by assuming an overlapping time coincidence window of $[-2, 2.5] \ \rm{h}$\footnote{The requirement for coincidence is a window of $[-2, 24] \ \rm{h}$ in \cite{moroianu2022assessment}.}.

\begin{figure}
    \hspace{-0.6cm}
    \includegraphics[scale=0.18]{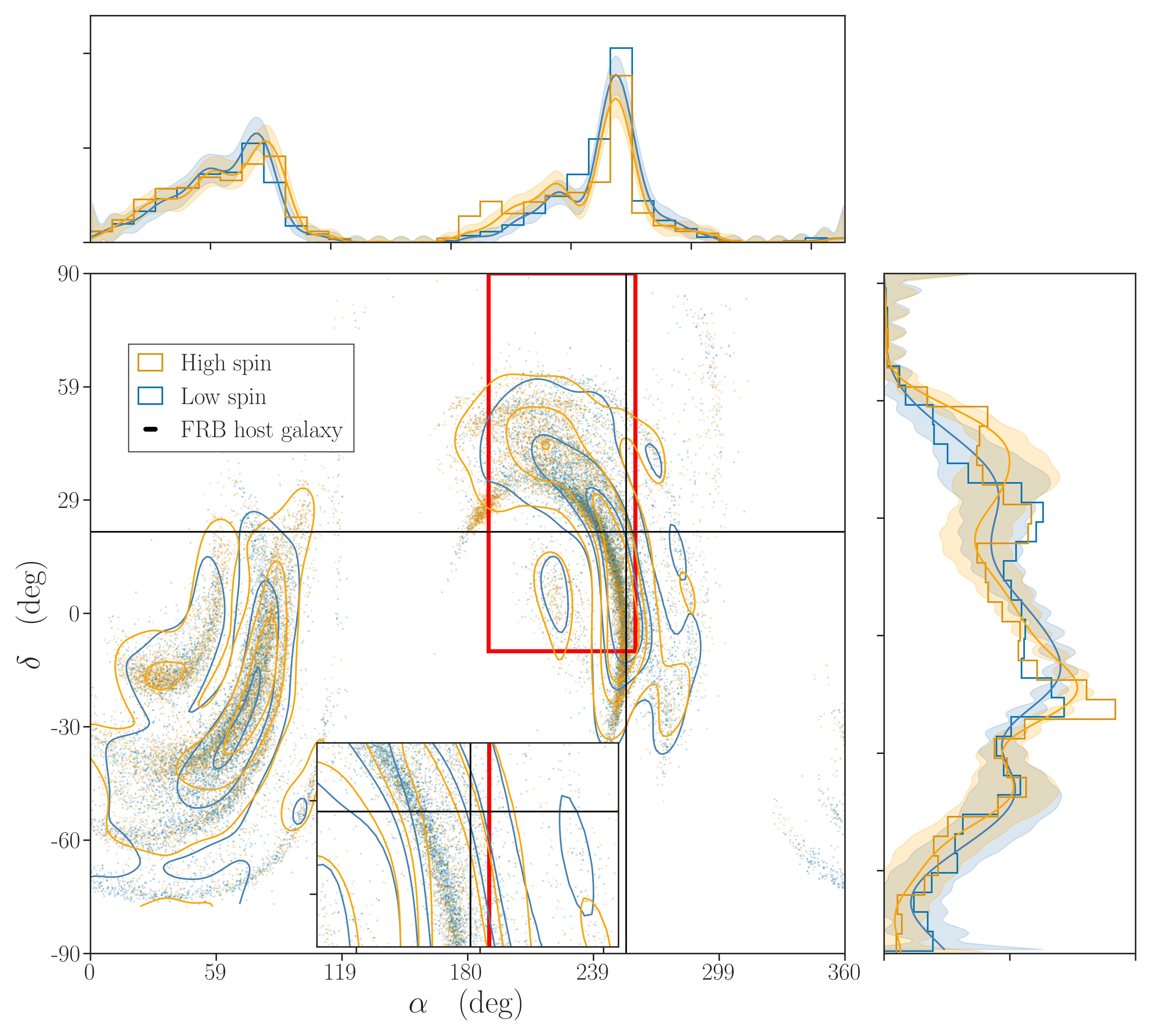}
    \caption{Skymap for \GW event, obtained both with \high and \low priors. The red box shows CHIME's instantaneous FOV at the time of \GW. Interpolation was generated with a two-dimensional GP density estimator (contour lines). The location of \FRB's most probable host galaxy (\GAL) is annotated for comparison.}
    \label{fig:skymap-GP}
\end{figure}

The instantaneous FOV of CHIME at the time of the event, centered around our choice of the overlapping time coincidence window defines $\pi(\Omega | \rm{FOV})$. This is shown as the red box in Figure~\ref{fig:skymap-GP}, it effectively restricts the full sky prior and hence we expect to reduce the chances for coincidence by about a factor of 5. The limited sky prior coincides with a large region of the \GW's skymap, where we have used a GP estimator to interpolate the LVK public samples.
We also note that the location of the presumed host galaxy falls just about within the $50\%$ probability contours of the GP density estimator, as shown by black crossing lines.

The posterior overlap integral results are shown in Table~\ref{table:odds-1} for both \low and \high samples and for both spatial priors $\pi(\Omega | \rm{FOV})$ and $\pi(\Omega|\rm{Full \ Sky})$. We note that the latter are only included for comparison, since we do believe that the full sky prior causes an overestimate of the association odds in the case of CHIME-detected counterparts.
The GP's \high results are consistent, within their uncertainty, with the values obtained with KDEs. 
The GP's \low results present a larger discrepancy with the KDEs, in some cases.
We blame this on the posterior surface of \low results being narrower and therefore the FRB location lying right on the edge of the 50\% spatial overlap probability contours. We also note that for the approximately disjoint integral $\mathcal{I}_{D_L}\mathcal{I}_{\Omega}$, most of the support is given by correlations with the luminosity distance, which increases the overall integral value.
We conclude the values obtained with the \high prior samples, which allow for the spinning HMNS hypothesis, are the most trustworthy, and hence we take $\mathcal{I}_{D_{L},\Omega}\approx10$.

\begin{deluxetable*}{cc|cccccccc}
    \tablecaption{Spatial overlap probabilities and constituent elements for two spin priors, calculated assuming Planck 2015 cosmology and using GWTC-3 samples. We report values obtained with two KDE methods (public LIGO \texttt{FITS} file for $\mathcal{I}_{\Omega}$ and \texttt{LIGO.skymap}'s \textit{ClusteredKDE} for $\mathcal{I}_{D_{\rm{L}},\Omega}$) and a Gaussian Process (GP) density estimator.}
    \label{table:odds-1}
    \tablehead{
    \multicolumn{2}{c}{} & \multicolumn{2}{c}{$\mathcal{I}_{D_{L}, \Omega}$} & \multicolumn{2}{c}{$\mathcal{I}_{D_{L}} \mathcal{I}_{\Omega}$} & \multicolumn{2}{c}{$\mathcal{I}_{D_{L}}$}  & \multicolumn{2}{c}{$\mathcal{I}_{\Omega}$} \\ \cline{3-10}
     \multicolumn{2}{c}{Prior assumptions} & \colhead{KDE} & \colhead{GP} & \colhead{KDE} & \colhead{GP} & \colhead{KDE} & \colhead{GP} & \colhead{KDE} & \colhead{GP}}
  \startdata
    Low-spin & $\pi(\Omega|\rm{Full \ Sky})$ & 45.7 & $72.5^{+4.6}_{-4.6}$ & 9.2 & $81.1^{+40}_{-40}$ & 12.4 & $12.7^{+6.1}_{-6.1}$ & 0.7 & $6.4^{+0.8}_{-0.8}$ \\
     & $\pi(\Omega|\textrm{FOV})$ & 8.9 & $14.1^{+0.9}_{-0.9}$ & 1.8 & $15.7^{+7.9}_{-7.9}$ & - & - & 0.1 & $1.2^{+0.1}_{-0.1}$ \\
    High-spin & $\pi(\Omega|\rm{Full \ Sky})$ & 52.1 & $50.3^{+3.8}_{-3.8}$ & 52.1 & $63.8^{+30}_{-30}$ & 13.4 & $13.5^{+6.1}_{-6.1}$ & 3.8 & $4.7^{+0.8}_{-0.8}$ \\
    & $\pi(\Omega|\textrm{FOV})$ & 10.1 & $9.8^{+0.7}_{-0.7}$ & 10.1 & $12.4^{+5.0}_{-5.0}$ & - & - & 0.7 & $0.9^{+0.1}_{-0.1}$ \\
    \enddata
\end{deluxetable*}

\subsection{Temporal overlap and prior odds}
Following \cite{ashton2018coincident}, we can write the temporal overlap integral for the time of coalescence $t_c$ for \GW and \FRB as, 
\begin{equation}
    \mathcal{I}_{t_c} = \begin{cases}
  \frac{T}{\Delta t}  & \text{if} \ (t_c - t_{\rm{EM}}) \in [\Delta t^{\rm{min}},\Delta t^{\rm{max}}]  \\
  0 & \text{otherwise}
\end{cases}
\end{equation}
where $\Delta t$ is defined as the window used to search for GW and FRB coincident events and where $T$ is the total co-observation time for both transient surveys. Now, the prior odds can be written in terms of the GW, EM, and joint detection rates as, 
\begin{equation}
    \pi_{C/R} \approx \frac{R_{\rm{GW,EM}}}{R_{\rm{GW}}R_{\rm{EM}}T}.
\end{equation}

Since we have little information on the rates of BNS detections with or without FRB counterparts, specifically FRB signals detectable by CHIME.
For the special case in which we are in $R_{\rm{GW}} \approx R_{\rm{GW,EM}} \ll R_{\rm{EM}}$, hence we must thus have,
\begin{equation}
    \pi_{C/R} \approx \frac{1}{R_{\rm{EM}}T}.
\end{equation}

\subsection{Posterior Odds}
We can now write the posterior odds between the coincident hypothesis $C$ and the random association $R$ by combining the spatial and temporal overlap integrals with the prior odds $\pi_{C/R}$. This choice leads to the posterior odds not to explicitly depend on the co-observation time $T$, 
we can therefore write the odds as, 
\begin{equation}
    \mathcal{O}_{C/R} \approx \frac{1}{R_{\rm{EM}}\Delta t} \mathcal{I}_{D_{L}, \Omega}
\end{equation}
We proceed to estimate $R_{\rm{EM}}$ by using the observed CHIME FRB detection rate using the latest catalog release \citep{CHIMEFRB:2021srp}. Using the 536 FRBs observed in 341 days, we estimate $R_{\rm{CHIME}} \approx 1.6 \ \rm{day}^{-1}$, where we have made the simplifying assumption that the CHIME instrument had zero downtime. 

The analysis performed in \cite{moroianu2022assessment} used a search window around O3a GW triggers of $\Delta t \approx 26 \ \rm{hours}$ (2 hours in the past and 24 hours in the future). Using the same search window we obtain $(R_{\rm{EM}}\Delta t)^{-1} \approx 0.5$. Consequently, the posterior odds are $\mathcal{O}_{C/R} \approx 5$ assuming \high spin prior. 

We can compute an optimistic estimate for the posterior odds by using a search window of $\Delta t \approx 3 \ \rm{hours}$ (corresponding roughly to the time delay between \GW and \FRB), obtaining $(R_{\rm{EM}}\Delta t)^{-1} \approx 5$ and thus the corresponding posterior odds are $\mathcal{O}_{C/R} \approx 50$. Our full results are summarized in Table~\ref{table:odds}. We positively highlight the sensitivity of our calculations to our prior assumptions, suggesting the importance of careful consideration of the latter. Our optimistic ($\Delta t \approx 3 \ \rm{hours}$) vs agnostic ($\Delta t \approx 26 \ \rm{hours}$) priors on the time window result in about $\mathcal{O}(10)$ discrepancy in the odds. Similarly, the discrepancy in results between un-informed (full-sky prior $\pi(\Omega|\rm{Full \ Sky})$) and informed (spatial overlap $\pi(\Omega|\textrm{FOV})$) priors is about $\mathcal{O}(5)$.

\begin{deluxetable*}{cc|cccc}
    \tablecaption{Posterior odds $\mathcal{O}_{C/R} $ calculated using the overlap integral $\mathcal{I}_{D_{\rm{L}}, \Omega}$ for two values for $\Delta t$: the actual search window used by Morianu et al (2022) and the approximate time delay between transients.
}
    \label{table:odds}
    \tablehead{
        \multicolumn{2}{c}{} & \multicolumn{2}{c}{$\Delta t \approx 26 \ \rm{hours}$} & \multicolumn{2}{c}{$\Delta t \approx 3 \ \rm{hours}$}\\
        \multicolumn{2}{c}{Prior assumptions} & KDE & GP & KDE & GP}
    \startdata
        Low-spin & $\pi(\Omega|\rm{Full \ Sky})$ &  22.8 & $36.2^{+2.3}_{-2.3}$ & 228 & $362^{+23}_{-23}$\\
         & $\pi(\Omega|\textrm{FOV})$ & 4.5 & $7.0^{+1.1}_{-1.1}$ & 44.5 & $70.5^{+4.5}_{-4.5}$\\
        High-spin & $\pi(\Omega|\rm{Full \ Sky})$ & 26.0 & $25.1^{+1.9}_{-1.9}$ & 260 & $251^{+19}_{-19}$\\
         & $\pi(\Omega|\textrm{FOV})$ & 5.0 & $4.9^{+0.3}_{-0.3}$ & 50.5 & $49^{+3.5}_{-3.5}$\\
     \enddata
\end{deluxetable*}

\section{\GW Parameter estimation with \GAL as its host galaxy} \label{section:pe}

We perform Bayesian parameter estimation with \texttt{BILBY} \citep{Ashton:2018jfp, Romero-Shaw:2020owr} using the \texttt{DYNESTY} nested sampling library \citep{Speagle:2019ivv}. We use the publicly available strain data for \GW \citep{strain-data-doi-lvk} observed by both the LIGO Hanford and Virgo detectors. To reduce the computational costs, the analysis is performed using the reduced order quadrature approximation \citep{Smith:2016qas, amanda_baylor_2019_3478659}, using the GW waveform model \texttt{IMRPhenomPv2\_NRTidal} \citep{Dietrich:2017aum, Dietrich:2018uni} which includes both tidal and precession effects as in \citep{LIGOScientific:2021usb}. We closely follow the analysis configuration performed in \citep{LIGOScientific:2021usb}, namely we maintain the same prior probability distributions on the GW binary parameters, such that we produce two sets of results: \low and \high priors. These correspond to dimensionless spin magnitudes for both components to be within the ranges $\chi_{1,2}<0.05$ and $\chi_{1,2}<0.89$ respectively.

To investigate the effects on the \GW parameter estimation results when we assume that \GAL was indeed the true host galaxy, we impose two distinct, and progressively stricter, constraints: we fix the sky location to $(\alpha, \delta)=(255.72^\circ, 21.52^\circ)$, the sky location of \GAL \citep{moroianu2022assessment}; and then we also fix the GW luminosity distance to the one of \GAL, corresponding to the spectroscopic redshift estimate for \GAL of $z =  0.03136 \pm 0.00009$ \citep{SDSS:2008tqn}, such that the galaxy position is fixed.

\begin{figure*}
\includegraphics[width=0.45\textwidth]{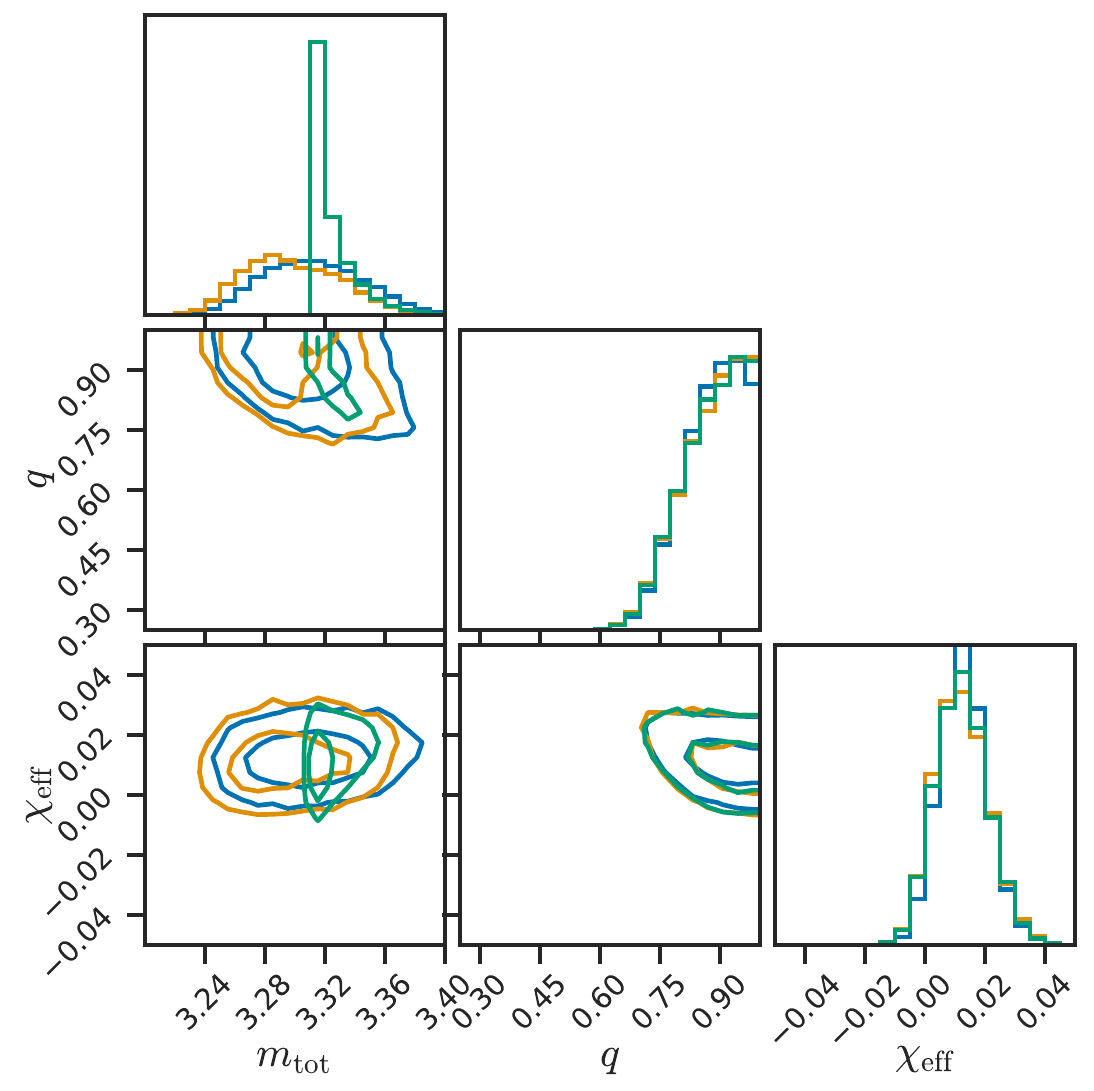}
\includegraphics[width=0.45\textwidth]{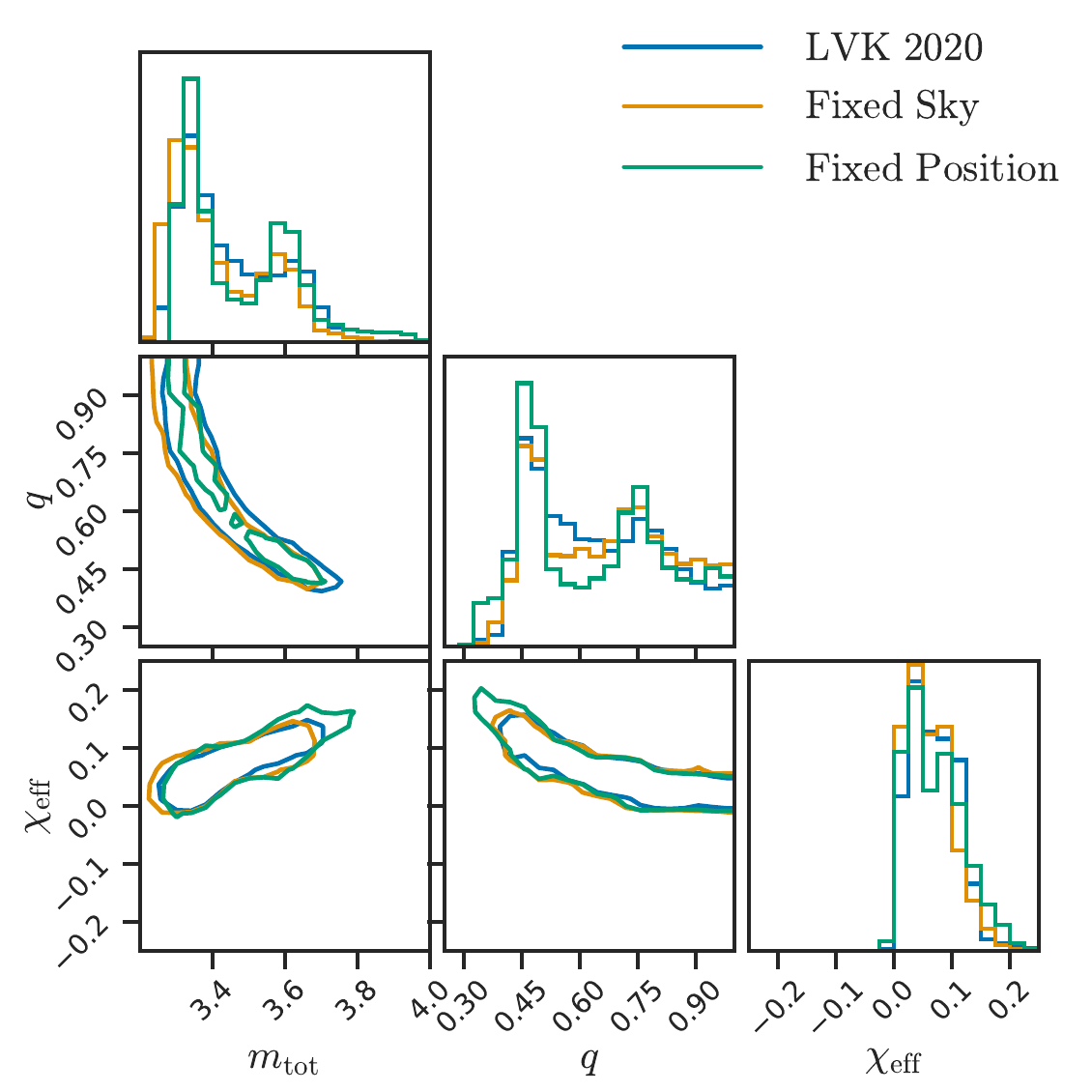}
\caption{\label{fig:corner} Corner plot showing the marginalized posterior distributions on the total mass $m_{\rm{tot}}$, the mass ratio $q$ and the effective inspiral spin parameter $\chi_{\rm{eff}}$ for \GW for both the low spin (left panel) and high spin (right panel) priors under the fixed sky location (orange) and fixed position (green) assumptions as described in \S\ref{section:pe}. For both cases, we also show the corresponding results from \cite{abbott2020gw190425} for reference (blue).}
\end{figure*}

\begin{figure*}
\includegraphics[width=0.45\textwidth]{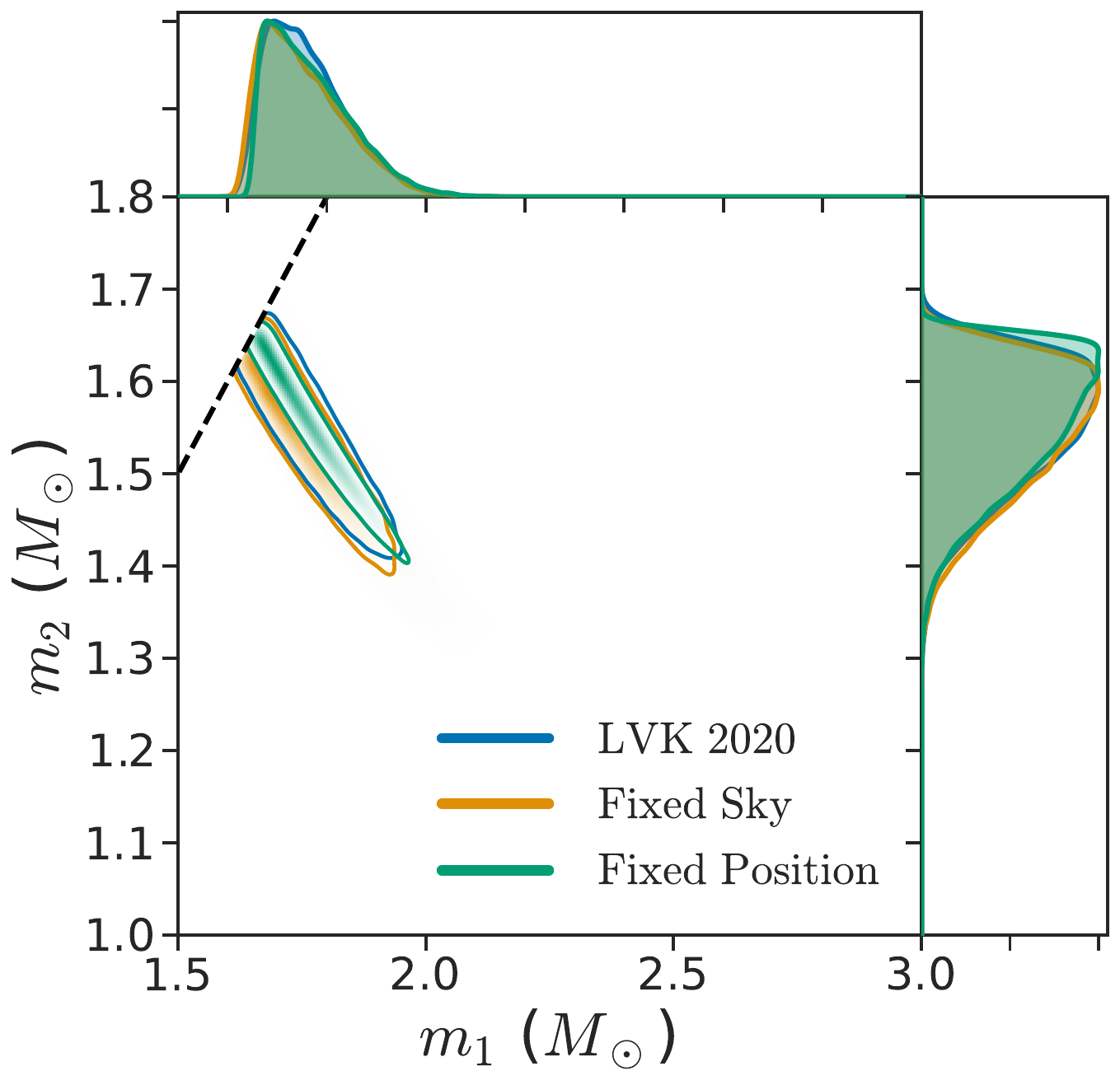}
\includegraphics[width=0.45\textwidth]{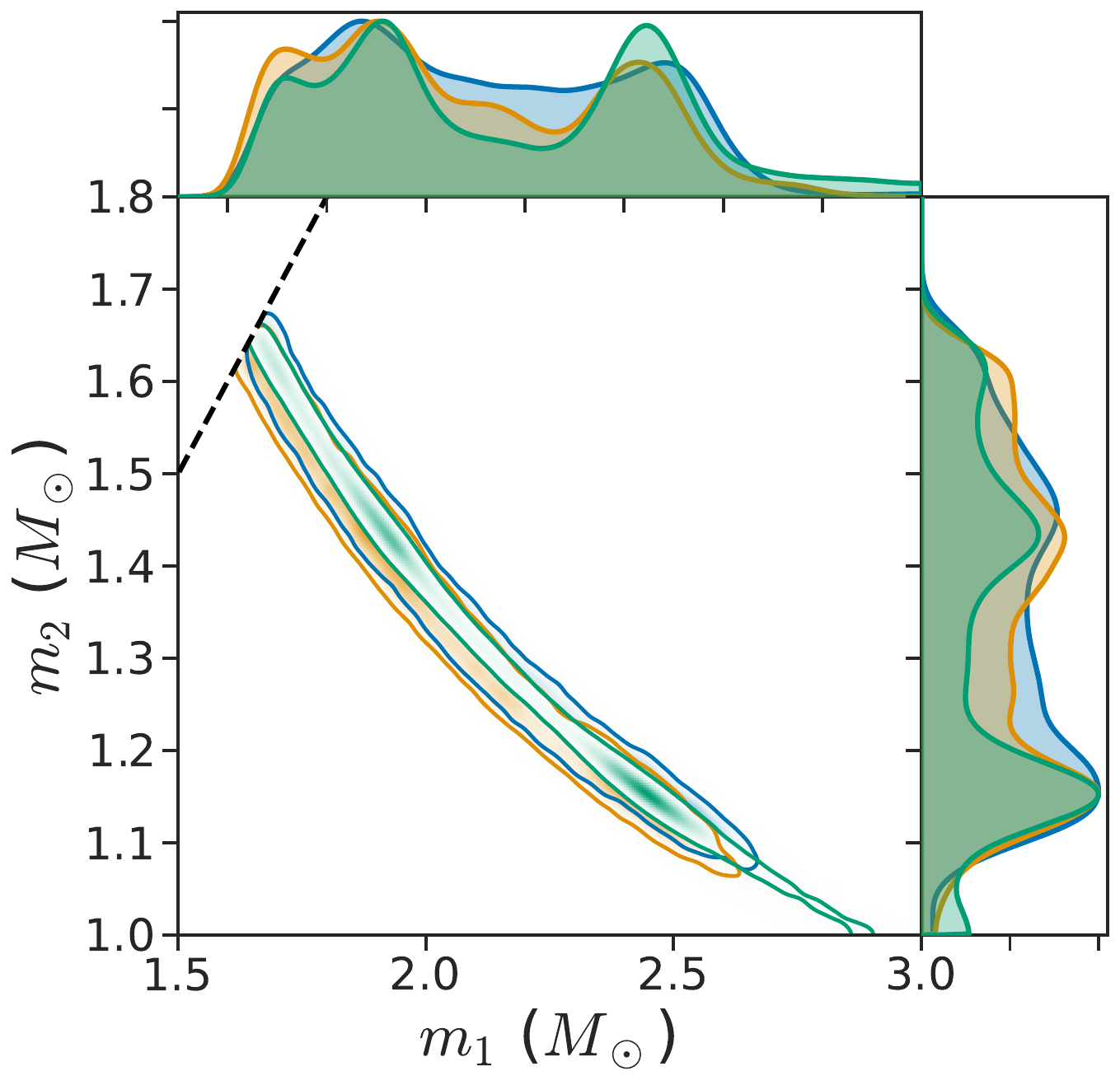}
\caption{\label{fig:m1m2} Marginalized posterior distributions on the primary mass $m_1$ and the secondary mass $m_2$ (both in the source frame) for \GW for both the low spin (left panel) and high spin (right panel) priors under the fixed sky location (orange) and fixed position (green) assumptions as described in \S\ref{section:pe}. For both cases, we also show the primary and secondary mass posteriors from \cite{abbott2020gw190425} for reference (blue).}
\end{figure*}

\begin{figure*}
\includegraphics[width=0.45\textwidth]{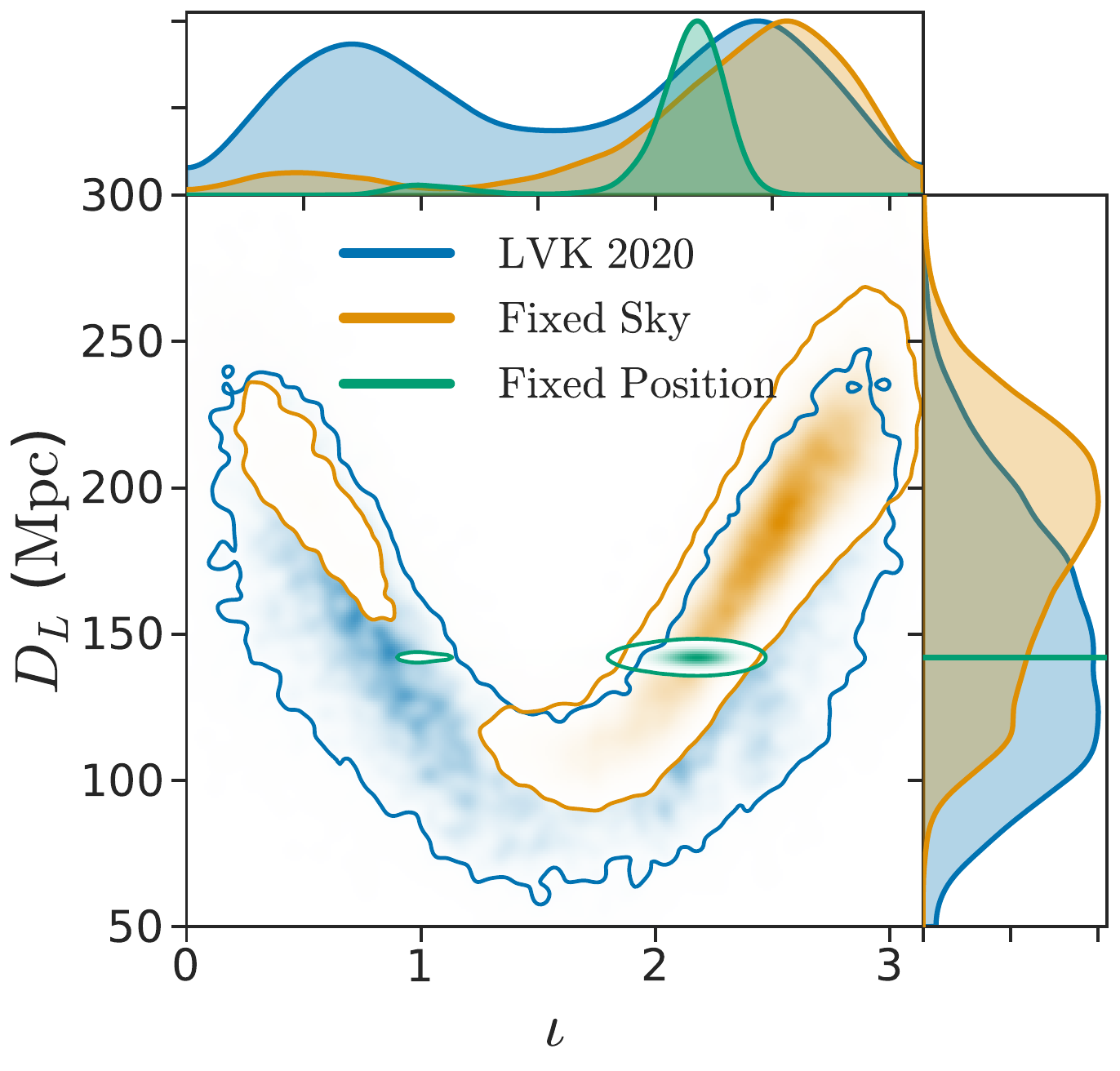}
\includegraphics[width=0.45\textwidth]{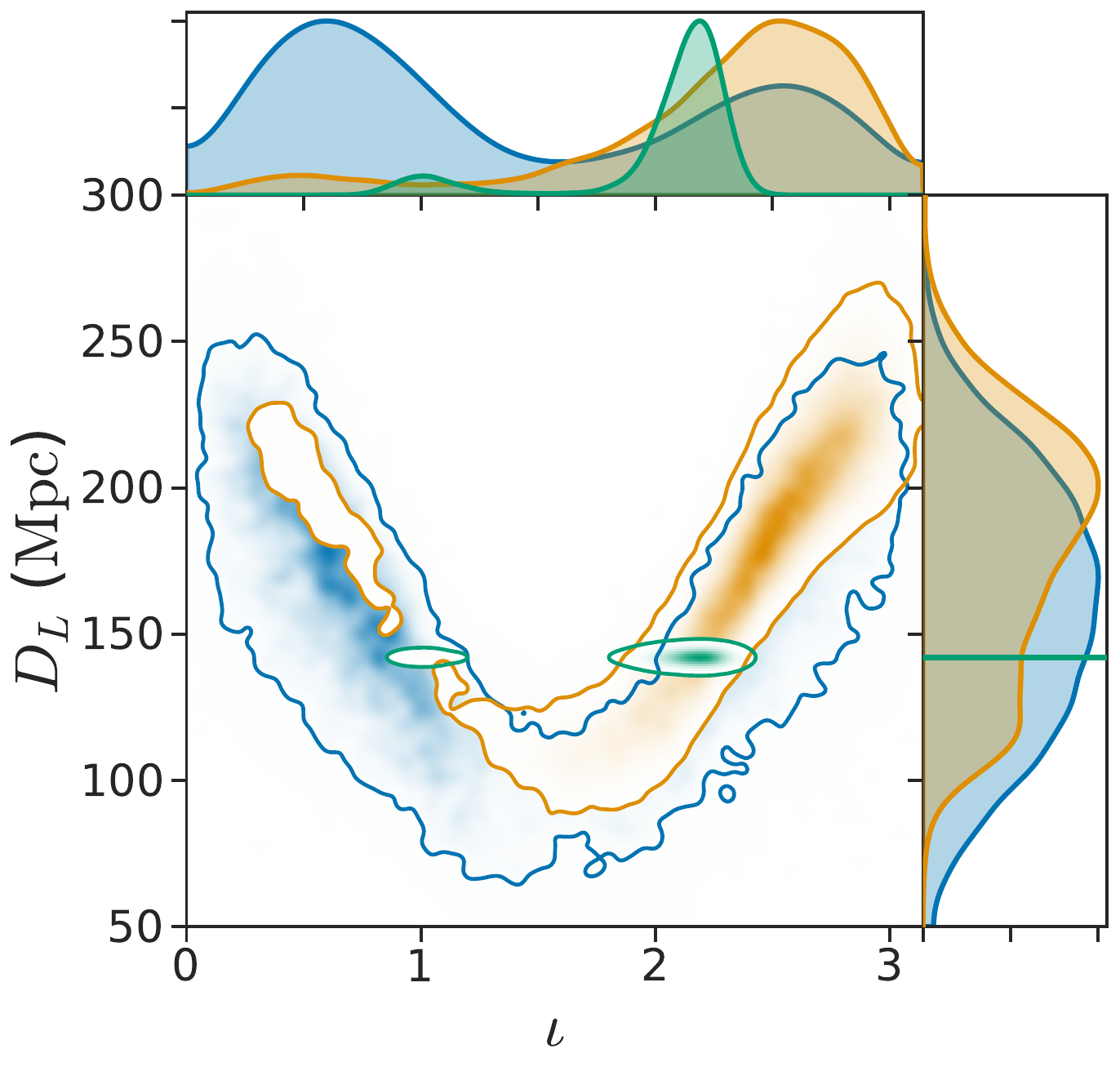}
\caption{\label{fig:dliota} Marginalized posterior distributions on the luminosity distance $D_L$ and the inclination angle $\iota$ for \GW for both the low spin (left panel) and high spin (right panel) priors under the fixed sky location (orange) and fixed position (green) assumptions as described in \S\ref{section:pe}. For both cases, we also show the results from \cite{abbott2020gw190425} for reference (blue).}
\end{figure*}

In Figure~\ref{fig:corner}, we show the inferred posterior distributions on the total mass $m_{\rm{tot}}$, the mass ratio $q$, and the effective inspiral spin parameter $\chi_{\rm{eff}}$ for \GW under both \low and \high assumptions, and with both the fixed sky and fixed position configurations. Similarly, in Figure~\ref{fig:m1m2}, we show the marginalized posterior distributions on the primary $m_1$ and the secondary $m_2$ masses (both in the source frame) for \GW. Lastly, the posterior distributions on the luminosity distance $D_L$ and the inclination angle $\iota$ are shown in Figure~\ref{fig:dliota}. For all results, we show the \GW posterior distribution inferred by the LVK \citep{abbott2020gw190425} for reference.

For the \low prior we find that the total mass and mass ratio are consistent with the LVK results for the fixed sky case, namely we find that the mass ratio must be greater than $q = 0.7$ at $68\%$ confidence and the total mass is $3.30^{+0.06}_{-0.04} \ M_{\odot}$. Meanwhile, when we fix the redshift to that of the \GAL galaxy position, the fixed position case only allows total masses greater than $m_\mathrm{tot} = 3.3 \ M_{\odot}$ at $99.9\%$ confidence, namely we find $3.32^{+0.04}_{-0.01} \ M_{\odot}$. For all runs, we find consistent posteriors on $\chi_\mathrm{eff}$. 

For the high spin prior we find consistent results for all intrinsic parameters. However, the mass ratio posteriors are bimodal allowing for mass ratios around $q = 0.45$ and we can constrain $q$ to be as low as $q=0.3$ with $99\%$ confidence. The total mass in this case is allowed to be higher than in the LVK case, due to the increased spin support, see Table~\ref{table:parameters} for explicit values. As for $\chi_\mathrm{eff}$, we find that $\chi_\mathrm{eff}>0$ at $99\%$ confidence, meaning that the binary can be highly spinning with positively aligned spins.

We find that fixing the sky location constrains the luminosity distance and inclination to approximately the same distribution for both spin prior assumptions, as shown in Figure~\ref{fig:dliota}. This effect can be understood as coming from the antenna pattern response functions, which depend on the sky location, constraining how the GW signal power is divided between both GW polarization amplitudes. 

\begin{figure}
    \centering
    \includegraphics[width=0.45\textwidth]{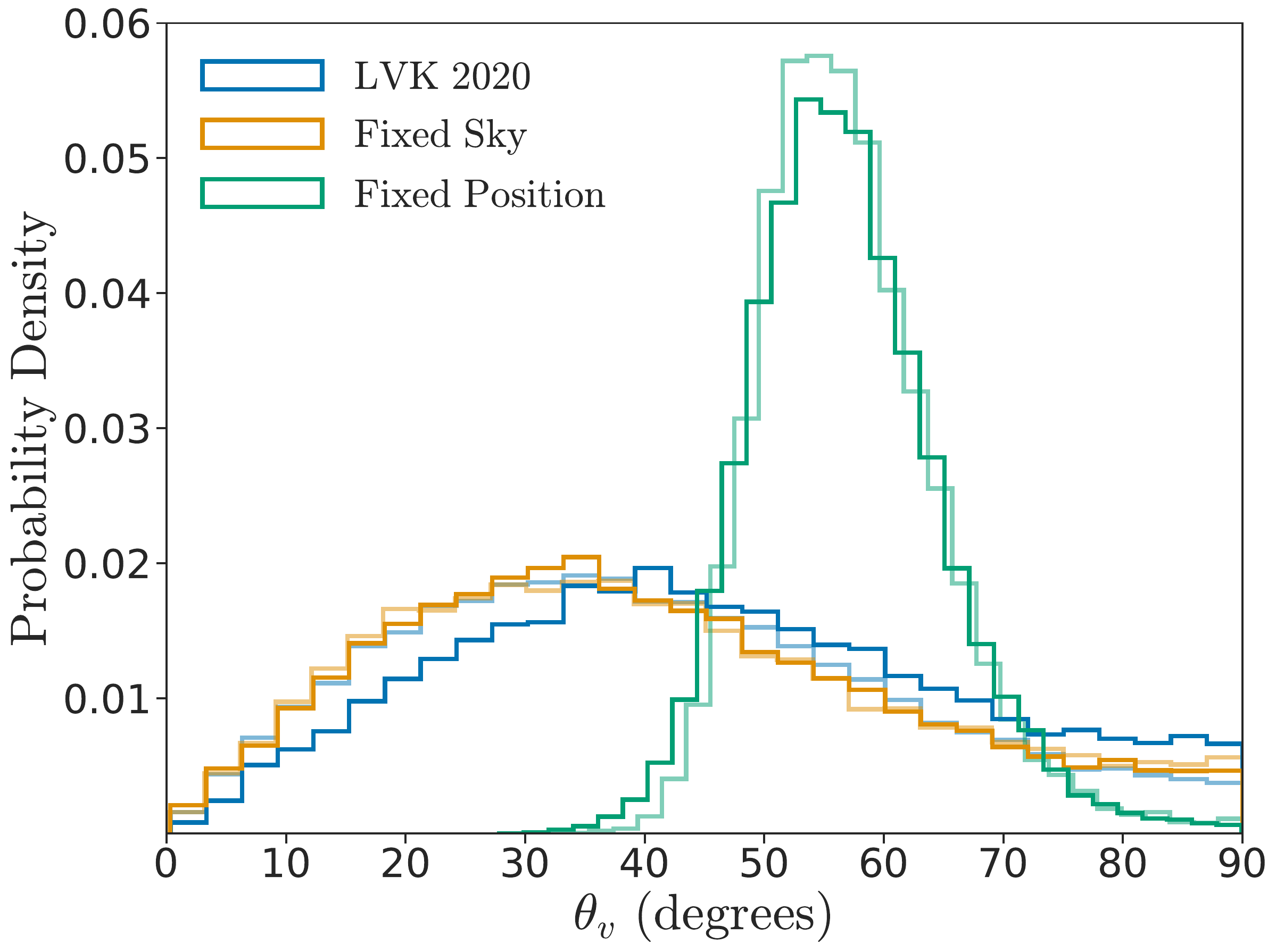}
    \caption{Marginalized posterior distributions on the viewing angle $\theta_v$ for \GW under the fixed sky location (orange) and fixed position assumptions (green) as described in \S\ref{section:pe}. For both cases, we also show the viewing angle posteriors computed using the results of \cite{abbott2020gw190425} for reference (blue). We show both the low spin (light lines) and high spin (solid lines) prior results for completeness.}\label{figure:viewing}
\end{figure}

Since the luminosity distance and inclination angle degeneracy is broken for all the assumptions considered in this work, it is useful to show the marginalized posterior distribution for the viewing angle $\theta_c$ (shown in Figure~\ref{figure:viewing}) to more clearly show that under the assumption that \GAL was indeed the host galaxy of \GW then it must have been an off-axis merger and consequently lead to effects on the expected EM emission \citep{Bhardwaj:2023avo}.

Finally, we provide a summary of the measured \GW parameters under the assumptions described in this section. The summary includes both the \low and \high prior results, with both the fixed sky and fixed position assumptions in Table \ref{table:parameters}.

\begin{deluxetable*}{lcccc} \label{table:parameters}
\tablecaption{Summary of updated parameters for \GW using both the \low and \high priors under the fixed sky and fixed position assumptions as described in Section \ref{section:pe}. We report all mass measurements in the source frame assuming a Planck 2015 cosmological model.
}
\tablehead{
    \colhead{~} & \multicolumn{2}{c}{Low-spin Prior} & \multicolumn{2}{c}{High-spin Prior}\\
    \colhead{~} & \colhead{Fixed Sky} & \colhead{Fixed Position} & \colhead{Fixed Sky} & \colhead{Fixed Position}
}
\startdata
Primary mass $m_1/M_{\odot}$ & $1.74^{+0.17}_{-0.09}$ & $1.75^{+0.17}_{-0.09}$ & $2.01^{+0.53}_{-0.33}$ & $2.10^{+0.59}_{-0.40}$\\
Secondary mass $m_2/M_{\odot}$ & $1.55^{+0.08}_{-0.14}$ & $1.57^{+0.08}_{-0.13}$ & $1.35^{+0.26}_{-0.25}$ & $1.32^{+0.30}_{-0.26}$\\
Chirp mass $\mathcal{M}/M_{\odot}$ & $1.43^{+0.02}_{-0.02}$ & $1.442^{+0.001}_{-0.001}$ & $1.43^{+0.02}_{-0.02}$ & $1.442^{+0.001}_{-0.001}$\\
Mass ratio $m_2/m_1$ & $0.89^{+0.10}_{-0.15}$ & $0.89^{+0.10}_{-0.15}$ & $0.67^{+0.29}_{-0.24}$ & $0.63^{+0.32}_{-0.24}$\\
Total mass $m_\mathrm{tot}/M_{\odot}$ & $3.30^{+0.06}_{-0.04}$ & $3.32^{+0.04}_{-0.01}$ & $3.37^{+0.28}_{-0.11}$ & $3.42^{+0.34}_{-0.11}$\\
Effective inspiral spin parameter $\chi_\mathrm{eff}$ & $0.01^{+0.02}_{-0.01}$ & $0.01^{+0.02}_{-0.01}$ & $0.06^{+0.08}_{-0.05}$ & $0.07^{+0.10}_{-0.06}$\\
Luminosity distance $D_\mathrm{L}$ & $183.7^{+58.2}_{-75.3} \ \rm{Mpc}$ & $-$ & $183.2^{+57.8}_{-73.3} \ \rm{Mpc}$ & $-$\\
Viewing angle $\theta_v$ & $37.8^{+42.4}_{-27.5} \ \rm{deg}$ & $56.1^{+14.3}_{-9.7} \ \rm{deg}$ & $37.8^{+41.3}_{-26.9} \ \rm{deg}$ & $55.6^{+14.3}_{-9.2} \ \rm{deg}$\\
\enddata
\end{deluxetable*}

\section{Conclusion} \label{section:conclusion}
In this work, we have investigated the association between the gravitational-wave event \GW and its presumed electromagnetic counterpart \FRB. We have re-calculated the probability of association as a Bayesian hypothesis comparing the hypothesis between a common source for the transients and the chance of a random source association. The posterior odds were calculated, following previous work, as a product of temporal and spatial overlap integrals. We found that the spatial overlap can marginally support a common source hypothesis, yielding a value of $\mathcal{O}(50)$. However, since both the CHIME observatory and the LIGO interferometers point in similar directions, the significance of the spatial overlap is lowered to $\mathcal{O}(10)$. The temporal overlap integrals yield less favorable results since they take into account the correlations between the instruments. The overall posterior odds were found to be $\mathcal{O}(5)$ and to only minimally support the association claimed by \cite{moroianu2022assessment}. 

We further investigate the association by re-running parameter estimation with the sky location of \GAL, the host galaxy for the \FRB counterpart identified by \cite{2023arXiv231010018B}, as well as with its measured redshift. The end-to-end parameter estimation analysis for the claimed associated transients is shown in this work in its entirety. Some of these, such as the viewing angle results, have been used by \cite{Bhardwaj:2023avo} to argue against the association hypothesis.

To conclude, we bring forward a word of caution when performing GW and EM counterpart associations, as shown in this work, simple spatial and temporal coincidences are useful and can in principle rule out potential associations (see \citealt{ashton2018coincident}), however, for the case considered, more observations of potentially associated GW and FRB counterparts will be needed to potentially shed light on the possibility of such transients having a common origin.

\section*{Acknowledgements} 
The authors would like to thank Jolien Creighton and Patrick Brady for their useful comments and feedback throughout this work. IMH is supported by a McWilliams postdoctoral fellowship at Carnegie Mellon University. IMH acknowledges support from NSF Award No. PHY-1912649 and PHY-2207728. VDE is supported by NSF's LIGO Laboratory which is a major facility fully funded by the National Science Foundation, operating under cooperative agreements PHY-1764464 and PHY-2309200. Additional support comes from NSF Award 2207758. The authors are grateful for computational resources provided by the Leonard E Parker Center for Gravitation, Cosmology and Astrophysics at the University of Wisconsin-Milwaukee and supported by NSF awards PHY-1912649, as well as computational resources provided by Cardiff University and supported by STFC grant ST/V001337/1 (UK LIGO Operations award). We thank LIGO and Virgo Collaboration for providing the data for this work. This research has made use of data, software and/or web tools obtained from the Gravitational Wave Open Science Center (https://www.gw-openscience.org/), a service of LIGO Laboratory, the LIGO Scientific Collaboration and the Virgo Collaboration. LIGO Laboratory and Advanced LIGO are funded by the United States National Science Foundation (NSF) as well as the Science and Technology Facilities Council (STFC) of the United Kingdom, the Max-Planck-Society (MPS), and the State of Niedersachsen/Germany for support of the construction of Advanced LIGO and construction and operation of the GEO600 detector. Additional support for Advanced LIGO was provided by the Australian Research Council. Virgo is funded, through the European Gravitational Observatory (EGO), by the French Centre National de Recherche Scientifique (CNRS), the Italian Istituto Nazionale di Fisica Nucleare (INFN) and the Dutch Nikhef, with contributions by institutions from Belgium, Germany, Greece, Hungary, Ireland, Japan, Monaco, Poland, Portugal, Spain. This material is based upon work supported by NSF's LIGO Laboratory which is a major facility fully funded by the National Science Foundation. This article has been assigned LIGO document number LIGO-P2400106.

\bibliography{main.bib}{}
\bibliographystyle{aasjournal}

\end{document}